\documentclass[PRB,noshowpacs,amsmath,amssymb,superscriptaddress,twocolumn]{revtex4}
\usepackage[T1]{fontenc}
\usepackage[latin1]{inputenc}
\usepackage{amsmath}
\usepackage{graphicx}
\usepackage{epsfig}
\usepackage{bm}
\usepackage{bbm}
\usepackage{dsfont}
\usepackage{color}

\newcommand{\nn}{\nonumber}

\newcommand{\bea}{\begin{eqnarray}}
\newcommand{\ea}{\end{eqnarray}}
\newcommand{\beq}{\begin{equation}}
\newcommand{\eq}{\end{equation}}
\newcommand{\bc}{\begin{center}}
\newcommand{\ec}{\end{center}}
\newcommand{\dg}{\dagger}
\newcommand{\la}{\langle}
\newcommand{\ra}{\rangle}
\newcommand{\ov}{\overline}

\begin{document}

\title{Expansion dynamics in two-dimensional Bose-Hubbard 
lattices:\\ Bose-Einstein condensate and thermal cloud}

\author{Mauricio Trujillo-Martinez}
\affiliation{Physikalisches Institut and Bethe Center for Theoretical Physics,
  Universit\"at Bonn, Nussallee 12, 53115 Bonn, 
Germany}

\author{Anna Posazhennikova}
\email[Email: ]{anna.posazhennikova@uni-greifswald.de}

\affiliation{Institut f\"ur Physik, Universit\"at Greifswald, 17487 Greifswald, Germany}

\author{Johann Kroha}
\email[Email: ]{kroha@th.physik.uni-bonn.de}

\affiliation{Physikalisches Institut and Bethe Center for Theoretical Physics,
  Universit\"at Bonn, Nussallee 12, 53115 Bonn, 
Germany}

\date{15 March 2021}

\begin{abstract}
We study the temporal expansion of an ultracold Bose gas 
in two-dimensional, square optical lattices. The gas is described by the
Bose-Hubbard model deep in the superfluid regime, with initially all bosons  
condensed in the central site of the lattice.  We use the previously developed 
nonequilibrium propagator method for capturing the time evolution of an 
interacting bosonic system, where the many-body Hamiltonian is represented 
in an appropriate local basis and the corresponding field operators are
separated into the classical [Bose-Einstein condensate (BEC)] part and quantum mechanical fluctuations. 
After a quench, i.e., after a sudden switch of the lattice nearest-neighbor 
hopping, the expanding bosonic cloud separates spatially into a fast 
ballistic forerunner and a slowly expanding central part controlled by 
self-trapping. We show that the forerunner expansion is driven by the
coherent dynamics of the BEC and that its velocity is 
consistent with the Lieb-Robinson bound. For smaller lattices we analyze how 
quasiparticle collisions lead to enhanced condensate depletion and  
oscillation damping. 
\end{abstract}


\maketitle

\section{Introduction}
\label{sec:intro}

The Bose-Hubbard model has been in the focus of condensed-matter research since the seminal paper by Fisher \textit{et al.} \cite{Fisher1989}, which predicted a superfluid--Mott-localization transition, and especially after its experimental realization in ultracold-atom systems \cite{Greiner2002}. It is well known that systems of cold atoms encompass a number of impressive advantages over their solid-state counterparts, among which are the full system controllability, freedom from impurities, isolation from the environment, and the possibility to realize quantum quenches. 
A quantum quench, i.e., an abrupt change of one of the system's parameters, is
a controlled way to bring a system out of equilibrium
\cite{Polkovnikov2011,Schmiedmayer2015,Eisert2015} 
and to study its nonequilibrium dynamics and potential thermalization 
\cite{Eisert2016,Rigol2016}.

The temporal expansion of clouds of interacting bosonic atoms in optical 
lattices has been studied experimentally
\cite{Schneider2013,Reinhard2013,Weiss2015,Schneider2015}
and theoretically 
\cite{Rigol2011,Rigol2013,Weiss2013,Heidrich-Meisner2013,Boschi2014,Heidrich-Meisner2015,Rigol2017}
as a straightforward way of accessing the nonequilibrium dynamics of these 
systems. Most of these studies used a low-density,
incoherent or Mott insulating initial state and investigated 
various aspects of the time evolution, such as self-trapping effects 
slowing down the expansion \cite{Reinhard2013,Rigol2013,Weiss2013}, 
a bimodal structure of the expanding cloud
\cite{Schneider2013,Schneider2015,Rigol2011,Heidrich-Meisner2015}, 
and doublon dynamics \cite{Boschi2014} and the expulsion of single 
bosons in the halo of the cloud, termed the quantum distillation 
effect \cite{Weiss2015}. 
Specifically, in the  experiment \cite{Schneider2013} a Mott insulating 
core with unity filling was prepared in the center of a two-dimensional (2D) optical
lattice. By abruptly lowering the lattice depth in one or both directions, 
tunneling and correspondingly expansion of the boson gas was induced and 
studied in detail in dependence on the interaction strength $U$. 
It was found that both dimensionality and interaction play an important 
role in the nonequilibrium expansion dynamics. In particular, the expansion 
in a 2D lattice developed bimodal cloud shapes with slow dynamics in the
round, central part of the cloud surrounded by fast, square-shaped ballistic 
wings. The core expansion velocity slowed down with the growth of  interaction
and eventually saturated \cite{Schneider2013}. On the theoretical side,
expansion dynamics in more than one spatial dimension has so far been studied 
on the level of mean-field theories 
\cite{Rigol2011,Rigol2013,Shapiro2007}, that is, 
considering the dynamics of the Bose-Einstein condensate (BEC) 
and neglecting the influence of the thermal cloud or incoherent excitations. 
In particular, time-dependent Gutzwiller mean-field theory was employed to
analyze the expansion dynamics in 2D bosonic Mott-Hubbard lattices 
\cite{Rigol2011,Rigol2013,Atav2014}. It was
found that initially confined Mott-insulating states become coherent during
the expansion after removal of a confining potential \cite{Rigol2011}, and
initially superfluid states with small occupation numbers expand in a bimodal
way with a central, slowly expanding cloud surrounded by a rapidly expanding
low-density cloud of bosons \cite{Rigol2013}. The bosons expand ballistically
and fastest along the diagonals of the lattice. In these works, the physics 
behind the expansion behavior was attributed to the fact, that the central 
cloud consists predominantly of doublons, which tend to group together
\cite{Muth2012}, whereas the fast expanding part consists of 
monomers \cite{Rigol2013,Weiss2015}. 

However, previous studies of the quench dynamics of  
Bose-Josephson junctions (BJJs) \cite{Mauro2009,Mauro2015,Anna2016,Annalen2018,
Lappe2018} showed that incoherent fluctuations
become inevitably excited after the quench even in gapped systems, 
due to the finite spectral distribution involved in the temporal evolution.
The coupling of the BEC to these incoherent excitations plays an important 
role in the relaxation and thermalization dynamics 
\cite{Anna2016,Annalen2018,Lappe2018}. 
Exact numerical calculations of the expansion dynamics are possible 
only in one dimension \cite{Schneider2013,Kollath2012}, where true Bose-Einstein
condensation does not occur and, hence, a sharp 
distinction between BEC and
incoherent excitations is not possible in the time-dependent situation.

In the present article we investigate the regime of high initial
density. In this regime,
doublon dynamics will not play an important role, so the effect of 
interaction-induced self-trapping on the expansion dynamics
can be studied independently. We also 
analyze the dynamics of the coherent BEC and of the noncondensed particle 
cloud separately and study the condensate depletion induced by the 
nonequilibrium dynamics.
To tackle the problem of the expansion of the coupled system of BEC and 
incoherent fluctuations we therefore adopt the semianalytical formalism
developed earlier and applied for studies of nonequilibrium BJJs
\cite{Mauro2009,Anna2016,Annalen2018}. Our approach goes beyond mean-field
approximations and systematically takes into account incoherent excitations. 
It consists of coupled equations of motion for the classical
space-time-dependent BEC amplitudes and for the full Keldysh quantum propagators
of the noncondensate fluctuations. 
These equations are derived from a generating functional using the 
Keldysh-Kadanoff-Baym nonequilibrium approach 
\cite{Kadanoffbook,Sieberer2016,Annalen2018}, which leads to 
a hierarchy of conserving approximations, and are solved self-consistently. 
We focus on the expansion dynamics of weakly interacting bosons in two 
dimensions, deep in the superfluid phase. For large square lattices of size $21\times 21$, we evaluate 
the theory within the leading-order-in-$U$ conserving approximation, 
the Bogoliubov-Hartree-Fock (BHF) approximation, while for smaller system 
size ($3\times 3$) we use the second-order self-consistent 
approximation, thus taking inelastic relaxation processes into account. 
Unlike previous treatments \cite{Rigol2011,Rigol2013} we can consider
arbitrary large numbers of particles within our formalism. 
Note, however, that we do not attain the Mott localized phase since the
interactions are assumed to be weak.

As the main results we find that even for small interaction strength
but large values of the initial local occupation number the coherent 
part of the expanding cloud splits into two modes, a slowly expanding 
high-density central part and a fast expanding low-density pulse 
or ``forerunner.'' In this case of high occupation number, however,
the expansion of the high-density part
is not inhibited by doublon formation \cite{Rigol2013} but by the
interaction-induced self-trapping effect known from the oscillation dynamics of  
BJJs \cite{Smerzi1997,Albiez2005,Mauro2009}. The forerunner expands 
ballistically, typical for coherent waves, instead of
diffusively. Despite the different initial conditions, these results 
are in good agreement with experiments on 2D 
expansion of low-density Bose gases \cite{Schneider2013}. 
The forerunner's expansion speed is largest along the lattice diagonals 
where it obeys and reaches the Lieb-Robinson bound \cite{Lieb1972}. 
The latter states that there is a finite limit to the speed of 
information propagation in any quantum system. 

In addition, we are able to distinguish the dynamics of the incoherent 
excitation cloud from the coherent one. The density of incoherent 
excitations increases with the interaction strength $U$, as expected, but 
with increasing $U$ it ceases to participate in the forerunner. 
This confirms that the forerunner is a coherence phenomenon. 
Taking inelastic two-particle scattering into account for small lattices 
beyond the BHF approximation, we find that it leads to fast damping 
of density oscillations between the BEC and the incoherent fluctuations 
and to enhanced depletion of the condensate.

%
%
\begin{figure}[b!]
\begin{center}
\hspace*{0.2cm} \includegraphics[width=\columnwidth]{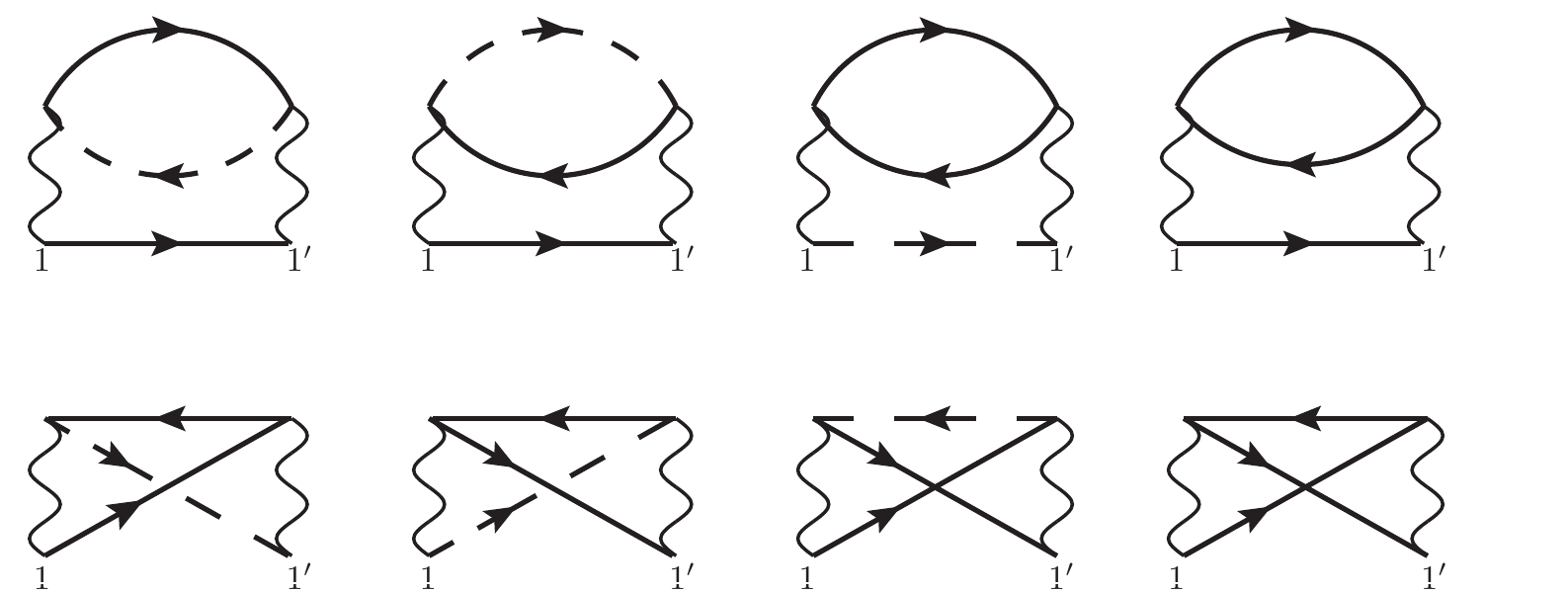} 
\end{center}
\caption{Diagrams contributing to the second-order self-energies. Indices $1$ and $1'$ refer to the first and second arguments of the self-energies ($1\equiv {\bf r},t$ and $1'\equiv {\bf r'},t'$). Solid lines represent propagators of noncondensed particles, whereas dashed lines denote propagators of condensed bosons. The wavy lines are associated with the interaction $U$. }
\label{fig:second_order}
\end{figure} 

This article is organized as follows.
In Sec. \ref{sec:model} we describe our model and derive the general kinetic
equations based on the Keldysh-Kadanoff-Baym nonequilibrium approach. 
Section \ref{sec:BHF} contains the equations of 
motion without two-particle damping, i.e., the dynamical BHF approximation, 
while in Sec. \ref{sec:second_order} the collision integrals describing 
inelastic interactions are taken into account. 
The results within the BHF approximation and including inelastic processes 
are presented and discussed in detail in
Secs. \ref{sec:resultsBHF} and \ref{sec:coll}, respectively. 
We summarize in Sec. \ref{sec:conclusion}. 

\onecolumngrid
\section{Model and equations of motion}
\label{sec:model}

\subsection{General kinetic equations}
\label{sec:gen_eqs}

We consider ultracold bosons in a 2D, square optical lattice described by the Bose-Hubbard Hamiltonian \cite{Jaksch1998}
\beq
H=-J(t)\sum_{\langle i,j \rangle}\hat b^{\dagger}_i\hat
b^{\phantom{\dagger}}_j+\sum_i \varepsilon_i\ \hat b^{\dagger}_i\hat b^{\phantom{\dagger}}_i+\frac{U}{2}\sum_i\hat n_i(\hat n_i-1),
\label{Bose_Hubbard}
\eq
where $U$ is the repulsive interaction between bosons, and
$J(t)=J\Theta(t)$ is the nearest neighbor hopping amplitude which is 
abruptly switched on at time $t=0$, with $\Theta(t)$ the Heaviside Theta 
function. In addition, $\hat b^{\dagger}_i$ and $\hat b_i$ are bosonic
creation and annihilation operators on
site $i$, respectively,
and $\hat n_i=\hat b^{\dagger}_i\hat b^{\phantom{\dagger}}_i$ is the
boson number operator. Hereafter we will consider a homogeneous lattice, 
$\varepsilon = \varepsilon_i$ for any $i$. 
As initial conditions at $t=0$ we assume all bosons condensed deep in the 
superfluid phase and trapped by a tight confining potential in the central 
site of the lattice. 

After the tunneling $J(t)$ is turned on for $t>0$, two main processes set in:
(i) the superfluid part is allowed to expand in all directions of the lattice
and (ii) incoherent excitations get excited and are expanding along with the
superfluid part. We analyze the resultant dynamics using the formalism
developed earlier \cite{Mauro2009,Mauro2015} based on standard nonequilibrium
techniques \cite{Kadanoffbook,Griffinbook,Rammerbook}. As usual, we decompose 
the bosonic operators into their expectation or saddlepoint value
$a_i(t)=\langle \hat b_i \rangle$ and the fluctuations $\hat \varphi_i$,  
\beq
\hat b_i= a_i(t)+\hat \varphi_i, \quad 
\hat b_i^{\dg}=a_i^*(t)+\hat \varphi_i^{\dg},
\eq
where $a_i(t)$ represents the local BEC amplitude and the noncondensate quantum
fluctuations are described by the operators $\hat \varphi_i=\hat b_i-a_i(t)$
obeying canonical bosonic commutation relations. We now treat the condensate
amplitudes semiclassically and the quantum fluctuations quantum
field theoretically in order to capture the nonequilibrium dynamics. 
The full bosonic propagator splits into two parts ${\bf C}+{\bf G}$, with 
\bea
{\bf C}_{ij}(t,t')&=&-{i}\left( \begin{matrix}
a_i(t)a_j^*(t') & a_i(t) a_j(t') \\
a_i^*(t)a_j^*(t')  & a_i^*(t) a_j(t')
\end{matrix}
\right) ,
\label{cond}\\
{\bf G}_{ij}(t,t')&=&-{i}\left( \begin{matrix}
\la T_C\hat{\varphi}_i(t)\hat{\varphi}_j^\dg (t') \ra & \la T_C\hat{\varphi}_i(t)\hat{\varphi}_j (t') \ra \\ 
\la T_C\hat{\varphi}_i^\dg(t)\hat{\varphi}_j^\dg (t') \ra  & \la T_C\hat{\varphi}_i^\dg(t)\hat{\varphi}_j (t') \ra
\end{matrix}
\right)  
=\left(\begin{matrix} G_{ij}(t,t' ) & F_{ij}(t,t' ) \\ \ov{F}_{ij}(t,t' ) & \ov{G}_{ij}(t,t' )   \end{matrix} \right),
\label{GF}
\ea
where $T_C$ denotes time-ordering operator along the Keldysh contour and 
${i}$ the imaginary unit. 
The general Dyson equations for the propagators ${\bf C}$ and ${\bf G}$ read
\bea
\sum_k \int\limits_C d \ov{t} \left[{\bf G }_{0,ik} ^{-1} (t,\ov{t}) - {\bf S}^{HF} _{ik}(t,\ov{t}) \right] {\bf C}_{kj}(\ov{t} , t'  ) 
 &=& \sum_k\int\limits_C d \ov{t}  {\bf S}_{ik}(t,\ov{t}){\bf C}_{kj}
(\ov{t} , t'), \label{dyson_eq1} \\
\sum_k \int\limits_C d \ov{t} \left[{\bf G }_{0,ik} ^{-1} (t,\ov{t}) - {\bf \Sigma}^{HF} _{ik}(t,\ov{t}) \right] {\bf G}_{kj}(\ov{t} , t'  ) 
 &=& \mathds{1} \delta(t-t')\delta_{ij}+\sum_k\int\limits_C d\ov{t}{\bf \Sigma}_{ik}(t,\ov{t}){\bf G}_{kj}(\ov{t},t').
\label{dyson_eq2}  
\ea
Here all integrals as well as the Dirac $\delta (t-t')$ are defined along
the Keldysh contour $C$. For convenience, we also separate the selfenergy
into two parts: the local Hartree-Fock (HF) part and the collision part,
described by ${\bf S}$ and ${\bf \Sigma}$, respectively \cite{Mauro2015,Annalen2018}. The inverse bare propagator reads
\beq
{\bf G}^{-1}_{0,ij}(t,t')=\left[{i}\,{\bf \boldsymbol{\tau}}_3\,\delta_{ij}\frac{\partial}{\partial t}+J\,\mathds{1}\, \delta_{\langle i,j\rangle} \right]\delta(t-t'),
\eq
where $\mathds{1}$ and $\boldsymbol{\tau}_3$ denote the unit and the third 
Pauli matrix in Bogoliubov particle-hole space, respectively. In addition,
$\delta_{\langle i,j\rangle}$ is $1$ if $i$ and $j$ are nearest neighbors
and zero otherwise. 
For further analysis it is convenient to rewrite the general Dyson
equations \eqref{dyson_eq1} and \eqref{dyson_eq2} in terms of symmetrized and antisymmetrized field-correlation functions
\bea
{\bf A}_{ij}(t,t')&=&{i}\left[ {\bf G}_{ij}^>(t,t')- {\bf G}_{ij}^<(t,t')\right]=\begin{pmatrix} \text{A}_{ij}^G & \text{A}_{ij}^F \\ \text{A}_{ij}^{\ov F} & \text{A}_{ij}^{\ov{G}}\end{pmatrix}, \nn \\
{\bf F}_{ij}(t,t')&=&\frac{1}{2}\left[ {\bf G}_{ij}^>(t,t')+ {\bf G}_{ij}^<(t,t')\right]=\begin{pmatrix} \text{F}_{ij}^G & \text{F}_{ij}^F \\ \text{F}_{ij}^{\ov F} & \text{F}_{ij}^{\ov{G}}\end{pmatrix}, \nn \\
\ea 
where we omitted the time arguments in the matrix representations for simplicity. Similarly, we introduce, for the self-energies,
\bea
{\bm\gamma}_{ij}(t,t')&=&{i}\left[ {\bf S}_{ij}^>(t,t')- {\bf S}_{ij}^<(t,t')\right]=\begin{pmatrix} \gamma_{ij}^G & \gamma_{ij}^F \\ \gamma_{ij}^{\ov F} & \gamma_{ij}^{\ov{G}}\end{pmatrix}, \nn \\
{\bf \Gamma}_{ij}(t,t')&=&{i}\left[ {\bf \Sigma}_{ij}^>(t,t')- {\bf \Sigma}_{ij}^<(t,t')\right]=\begin{pmatrix} \Gamma_{ij}^G & \Gamma_{ij}^F \\ \Gamma_{ij}^{\ov F} & \Gamma_{ij}^{\ov{G}}\end{pmatrix}, \nn \\
{\bf \Pi}_{ij}(t,t')&=&\frac{1}{2}\left[ {\bf \Sigma}_{ij}^>(t,t')+ {\bf \Sigma}_{ij}^<(t,t')\right]=\begin{pmatrix} \Pi_{ij}^G & \Pi_{ij}^F \\ \Pi_{ij}^{\ov F} & \Pi_{ij}^{\ov{G}}\end{pmatrix}. \nn \\
\ea
The symbols $<$ and $>$ refer to the standard lesser and greater notations for nonequilibrium Green's functions and self-energies \cite{Rammerbook}. 

With this new notation we can rewrite Eqs. \eqref{dyson_eq1} and \eqref{dyson_eq2} in terms of symmetrized and antisymmetrized correlators and their selfenergies
\bea
{i}\boldsymbol{\tau}_3\frac{\partial}{\partial t}{\bf C}_{ij}&=&-J(\sum_{i':\langle i',i\rangle}{\bf C}_{i',j}+\sum_{j':\langle j',j\rangle}{\bf C}_{i,j'})+{\bf S}_i^{HF}{\bf C}_{ij}  -{i}\sum_k\int\limits_0^{t}d\ov{t}{\bm\gamma}_{ik}{\bf C}_{kj}, \label{general_eq1}\\
{i}\boldsymbol{\tau}_3\frac{\partial}{\partial t}{\bf A}_{ij}&=&-J(\sum_{i':\langle i',i\rangle}{\bf
  A}_{i',j}+\sum_{j':\langle j',j\rangle}{\bf A}_{i,j'})+{\bf \Sigma}_i^{HF}{\bf A}_{ij}-{i}\sum_k\int\limits_{t'}^{t}d\ov{t}{\bf \Gamma}_{ik}{\bf
  A}_{kj}, \label{general_eq2} \\
{i}\boldsymbol{\tau}_3\frac{\partial}{\partial t}{\bf F}_{ij}&=&-J(\sum_{i':\langle i',i\rangle}{\bf F}_{i',j}+\sum_{j':\langle j',j\rangle}{\bf F}_{i,j'})+{\bf \Sigma}_i^{HF}{\bf F}_{ij} -\ {i}\sum_k\int\limits_0^{t}d\ov{t}{\bf \Gamma}_{ik}{\bf F}_{kj}+{i}\sum_k\int\limits_{0}^{t'}d\ov{t}{\bf \Pi}_{ik}{\bf A}_{kj}.
\label{general_eq3}
\ea
In Eqs.~(\ref{general_eq2}) and (\ref{general_eq3}) we used the locality of the 
Hartree-Fock selfenergies in the time arguments as well as in the site
indices, i.e., ${\bf S}_{ij}^{HF}={\bf
  S}_i^{HF}\delta_{ij}\delta(t-t')$ and similarly for ${\bf \Sigma}^{HF}_{ij}$,
and evaluated the contour integrals. Here and in the following, a
  sum index $i':\langle i',i\rangle$ indicates summation of $i'$ over nearest neighbors of fixed $i$. Because of the quench boundary conditions, the integrals involving collisional self-energies
in Eqs. \eqref{general_eq1}--\eqref{general_eq3} start from $0$ and not from $-\infty$. For
simplicity, we omitted again explicit time arguments of the Green's functions
and self-energies. The argument $t$ refers to the first argument of the
functions and selfenergies, that is, ${\bf C}_{ij}\equiv {\bf C}_{ij}(t,t')$ and  $\int\limits_{t'}^{t}d\ov{t}\,{\bf \Gamma}_{ik}{\bf A}_{kj}=\int\limits_{t'}^{t}d\ov{t}\,{\bf \Gamma}_{ik}(t,\ov{t}){\bf A}_{kj}(\ov{t},t')$, etc. The boundary conditions are formulated in such a way that the system of equations \eqref{general_eq1}--\eqref{general_eq3} becomes an initial value problem.

\subsection{Bogoliubov-Hartree-Fock approximation}
\label{sec:BHF}

In this section we specify the set of differential equations derived from
\eqref{general_eq1}--\eqref{general_eq3} in the BHF approximation, i.e., the
first-order-in-$U$ conserving approximation \cite{Griffinbook}. To first order
in $U$ we 
neglect the integrals in Eqs. \eqref{general_eq1}--\eqref{general_eq3} and
explicitly calculate the HF self-energies ${\bf S}^{HF}$ and ${\bf
  \Sigma}^{HF}$ as
\bea
{\bf \Sigma}_i^{HF}&=&\begin{pmatrix} \Sigma_i^{HF} & \Omega_i^{HF} \\ \ov{\Omega}_i^{HF} & \ov{\Sigma}_i^{HF}\end{pmatrix} 
= {i}U\left[\left( \frac{1}{2}\text{Tr}[{\bf C}_{ii}]\mathds{1}+{\bf C}_{ii}\right)+\left( \frac{1}{2}\text{Tr}[{\bf F}_{ii}]\mathds{1}+{\bf F}_{ii}\right)\right],  \\
{\bf S}_i^{HF}&=&\begin{pmatrix} S_i^{HF} & W_i^{HF} \\ \ov{W}_i^{HF} & \ov{S}_i^{HF}\end{pmatrix} 
= {i}U\left[\frac{1}{2}\text{Tr}[{\bf C}_{ii}]\mathds{1}+\left( \frac{1}{2}\text{Tr}[{\bf F}_{ii}]\mathds{1}+{\bf F}_{ii}\right)\right].
\ea
Note that, in this approximation, quantities related to the spectral function ${\bf A}_{ij}$  decouple from the system of equations, and the final system of BHF equations reads
\begin{equation}
\begin{aligned}\relax
& {i}\frac{\partial }{\partial t}a_i=-J\sum_{i':\langle i',i\rangle} a_{i'}+(U|a_i|^2+2{i}U\text{F}_{ii}^G)a_i+{i}U\text{F}_{ii}^Fa_i^*,  \\
& {i}\frac{\partial }{\partial t}\text{F}^G_{ij}=-J\Bigl(\sum_{i':\langle i',i\rangle}\text{F}^G_{i',j}+\sum_{j':\langle j',j\rangle}\text{F}^G_{i,j'}\Bigr)  
 +\Sigma_i^{HF}\text{F}^F_{ij}-\text{F}^F_{ij}\Sigma_j^{HF}-\Omega_i^{HF}(\text{F}^F_{ij})^*-\text{F}^F_{ij}(\Omega_j^{HF})^*,  \\
 & {i}\frac{\partial }{\partial t}\text{F}^F_{ij}=-J\Bigl(\sum_{i':\langle i',i\rangle} \text{F}^F_{i',j}+\sum_{j':\langle j',j\rangle} \text{F}^F_{i,j'} \Bigr)
 +\Sigma_i^{HF}\text{F}^F_{ij}+\text{F}^F_{ij}\Sigma_j^{HF}-\Omega_i^{HF}(\text{F}^G_{ij})^*-\text{F}^G_{ij}\Omega_j^{HF}.  
\end{aligned}
\label{HF_eq}
\end{equation}
We then solve the equations numerically, the results being discussed  in detail in section \ref{sec:resultsBHF}. 

\vspace*{0.5cm}

\subsection{Selfconsistent second-order approximation}
\label{sec:second_order}

In the self-consistent second-order approximation, accounting for collisions
and relaxation, the equations are
significantly  more complicate, and the spectral and statistical parts are
coupled, unlike in the BHF approximation. 
The diagrammatic contributions to the self-energies in second order are
depicted in Fig. \ref{fig:second_order}. Due to the underlying symmetry relations
specified in Appendix A, we give here only the expressions for the upper-left and
upper-right components of the second-order matrix self-energy ${\bm\gamma_{ij}}$ [see Eqs. \eqref{general_eq1}--\eqref{general_eq3}],
\bea
\gamma^G_{ij}&=&U^2\text{F}^G_{ij}(4\Lambda_{ij}[F,F^*]+2\Lambda_{ij}[G,G^*]) 
+U^2\text{A}^G_{ij}(4\Xi_{ij}[F,F^*]+2\Xi_{ij}[G,G^*]),  \\
\gamma^F_{ij}&=&U^2\text{F}^F_{ij}(4\Lambda_{ij}[G,G^*]+2\Lambda_{ij}[F,F^*])
+U^2\text{A}^F_{ij}(4\Xi_{ij}[G,G^*]+2\Xi_{ij}[F,F^*]).
\ea
Here we introduced the shorthand notation
\bea
\Lambda_{ij}[f,g](t,t')\hspace*{-0.1cm}&=&\hspace*{-0.1cm}\text{A}_{ij}^f(t,t')\text{F}_{ij}^g(t,t')+\text{A}_{ij}^g(t,t')\text{F}_{ij}^f(t,t'), \nn \\
\Xi_{ij}[f,g](t,t')\hspace*{-0.1cm}&=&\hspace*{-0.1cm}\text{F}_{ij}^f(t,t')\text{F}_{ij}^g(t,t')-\frac{1}{4}\text{A}_{ij}^f(t,t')\text{A}_{ij}^g(t,t'),\nn\\ 
\ea
where $g,f\in\{G,F,G^*,F^*\}$ and $A^{g^*}\equiv (A^g)^*$.
The other two contributions ${\bf \Gamma}_{ij}$ and ${\bf \Pi}_{ij}$ from Eqs. \eqref{general_eq1}--\eqref{general_eq3}
are expressed as 
\bea
\Gamma_{ij}^G=2{i}U^2(2a_i^*a_j^*\Lambda_{ij}[F,G]+a_i^*a_j\Lambda_{ij}[G,G]-2a_ia_j\Lambda_{ij}[F^*,G] 
-2a_ia_j^*\Lambda_{ij}[G,G^*]-2a_ia_j^*\Lambda_{ij}[F,F^*])  \nn \\
+U^2(\text{F}^G_{ij}\{4\Lambda_{ij}[F,F^*]+2\Lambda_{ij}[G,G^*]\} 
+\text{A}_{ij}^G\{4\Xi_{ij}[F,F^*]+2\Xi_{ij}[G,G^*]\}),  \\
\Gamma_{ij}^F=2{i}U^2(2a_i^*a_j\Lambda_{ij}[F,G]+a_i^*a_j^*\Lambda_{ij}[F,F]-2a_ia_j^*\Lambda_{ij}[F,G^*] 
-2a_ia_j\Lambda_{ij}[G,G^*]-2a_ia_j\Lambda_{ij}[F,F^*]) \nn \\
+U^2(\text{F}^F_{ij}\{4\Lambda_{ij}[G,G^*]+2\Lambda_{ij}[F,F^*]\} 
+\text{A}_{ij}^F\{4\Xi_{ij}[G,G^*]+2\Xi_{ij}[F,F^*]\}),  
\ea
and
\bea
\Pi_{ij}^G=2{i}U^2(2a_i^*a_j^*\Xi_{ij}[F,G]+a_i^*a_j\Xi_{ij}[G,G]  
-2a_ia_j\Xi_{ij}[F^*,G]-2a_ia_j^*\Xi_{ij}[G,G^*]-2a_ia_j^*\Xi_{ij}[F,F^*])  \nn \\
+U^2(\text{F}^G_{ij}\{4\Xi_{ij}[F,F^*]+2\Xi_{ij}[G,G^*]\}  
-\frac{1}{2}\text{A}_{ij}^G\{2\Lambda_{ij}[F,F^*]+\Lambda_{ij}[G,G^*]\}),  \\
\Pi_{ij}^F=2{i}U^2(2a_i^*a_j\Xi_{ij}[F,G]+a_i^*a_j^*\Xi_{ij}[F,F]  
-2a_ia_j^*\Xi_{ij}[F,G^*]-2a_ia_j\Xi_{ij}[G,G^*]-2a_ia_j\Xi_{ij}[F,F^*])  \nn \\
+U^2(\text{F}^F_{ij}\{4\Xi_{ij}[G,G^*]+2\Xi_{ij}[F,F^*]\}  
-\frac{1}{2}\text{A}_{ij}^F\{2\Lambda_{ij}[G,G^*]+\Lambda_{ij}[F,F^*]\}).
\label{gamma_pi}
\ea
With these self-energies, the equations of motion for the spectral and 
for the statistical functions  become, respectively,
\bea
{i}\frac{\partial \text{A}^G_{ij}}{\partial t}=-J\Bigl(\sum_{i':\langle i',i\rangle} \text{A}_{i',j}^G+\sum_{j':\langle j',j\rangle} \text{A}^G_{i,j'}\Bigr)+\Sigma_i^{HF}\text{A}^G_{ij} 
-\Omega_i^{HF}(\text{A}^F_{ij})^*-{i}\sum_k\int \limits_{t'}^{t}d \ov{t}(\Gamma^G_{ik}\text{A}^G_{kj}+\Gamma^F_{ik}\text{A}^{\ov{F}}_{kj}),  \label{eq:spectral1} \\
{i}\frac{\partial \text{A}^F_{ij}}{\partial t}=-J\Bigl(\sum_{i':\langle i',i\rangle} \text{A}_{i',j}^F+\sum_{j':\langle j',j\rangle} \text{A}^F_{i,j'}\Bigr)+\Sigma_i^{HF}\text{A}^F_{ij} 
-\Omega_i^{HF}(\text{A}^G_{ij})^*-{i}\sum_k\int \limits_{t'}^{t}d \ov{t}(\Gamma^G_{ik}\text{A}^F_{kj}+\Gamma^F_{ik}\text{A}^{\ov{G}}_{kj}),
\label{eq:spectral2}
\ea
\bea
{i}\frac{\partial \text{F}^G_{ij}}{\partial t}\hspace*{-0.1cm}&=\hspace*{-0.1cm}&-J\Bigl(\sum_{i':\langle i',i\rangle} \text{F}_{i',j}^G+\sum_{j':\langle j',j\rangle} \text{F}^G_{i,j'}\Bigr)+\Sigma_i^{HF}\text{F}^G_{ij} 
-\Omega_i^{HF}(\text{F}^F_{ij})^*-{i}\sum_k\int \limits_{0}^{t}d \ov{t}(\Gamma^G_{ik}\text{F}^G_{kj+\Gamma^F_{ik}\text{F}^{\ov{F}}_{kj})} \label{eq:stat1}\\
&&\hspace*{8.4cm}+{i}\sum_k\int \limits_{0}^{t'}d
\ov{t}(\Pi^G_{ik}\text{A}^G_{kj}+\Pi^F_{ik}\text{A}^{\ov{F}}_{kj}),\nn\\  
{i}\frac{\partial \text{F}^F_{ij}}{\partial t}\hspace*{-0.1cm}&=&\hspace*{-0.1cm}-J\Bigl(\sum_{i':\langle i',i\rangle} \text{F}_{i',j}^F+\sum_{j':\langle j',j\rangle} \text{F}^F_{i,j'}\Bigr)+\Sigma_i^{HF}\text{F}^F_{ij} 
-\Omega_i^{HF}(F^G_{ij})^*-{i}\sum_k\int \limits_{0}^{t}d \ov{t}(\Gamma^G_{ik}\text{F}^F_{kj}+\Gamma^F_{ik}\text{F}^{\ov{G}}_{kj})  \label{eq:stat2}\\ 
&&\hspace*{8.3cm}+{i}\sum_k\int \limits_{0}^{t'}d \ov{t}(\Pi^G_{ik}\text{A}^F_{kj}+\Pi^F_{ik}\text{A}^{\ov{G}}_{kj}).\nn
\ea
Here $t$ refers again to the first time argument of the entities and $t'$ to
the second one. For the notation in the convolution integrals in 
these equations, see the end of Sec. \ref{sec:gen_eqs}. The sets of equations 
\eqref{eq:spectral1} and \eqref{eq:spectral2}, and
\eqref{eq:stat1} and \eqref{eq:stat2} are coupled to the equations for the
condensate amplitudes on sites $i$,
\bea
{i}\frac{\partial a_i}{\partial t}=-J\,\sum_{i':\langle i',i\rangle}a_{i'}+(U|a_i|^2+2{i}U\text{F}^G_{ii})a_i 
+{i}U\text{F}^F_{ii}a_i^*-{i}\sum_k\int\limits_{0}^{t}d\ov{t}(\gamma_{ik}^Ga_k+\gamma_{ik}^Fa_k^*). 
\label{eq:cond}
\ea
Note that the integrals in Eqs.\eqref{eq:spectral1}--\eqref{eq:stat2} still contain
$\ov{G}$ and $\ov{F}$ components of spectral and statistical functions. The
treatment of such integrals using the symmetry relations from Appendix
\ref{symmetry_rel} is described in detail in Appendix \ref{integrals}. 
\vspace*{0.8cm}
\twocolumngrid

%
%
\begin{figure*}[t!]
\vspace*{-0.4cm}
  \centering
\includegraphics[width=0.99\textwidth]{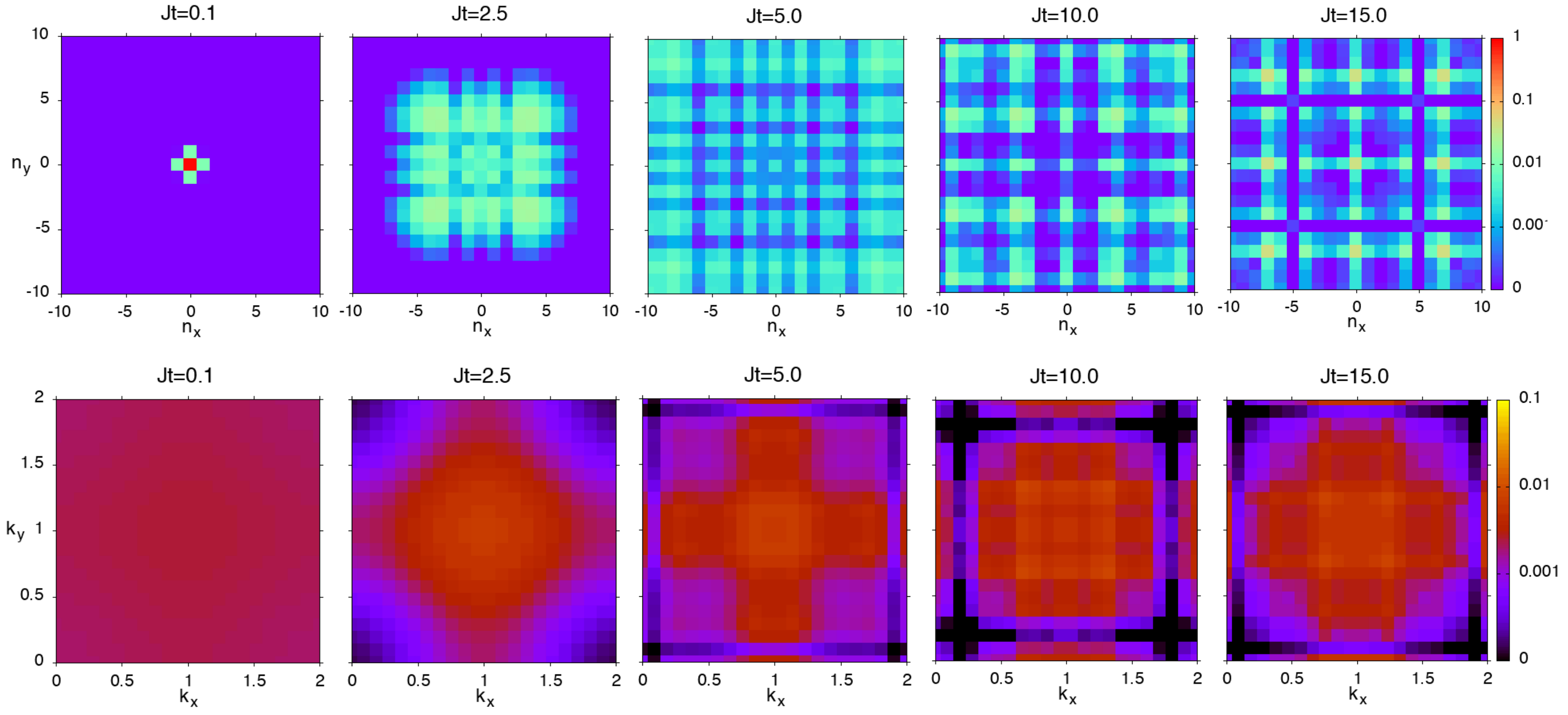}
\caption{(Color online) Expansion dynamics of the BEC cloud for interaction strength $u=3$, $\rho=1$, and lattice size $21\times 21$. The top row shows the expansion in real space for various times from $tJ=0.1$ to $15$. The color code represents the local condensate occupation numbers in units of the total particle number $N^0_i(t)/N$. The lattice momentum distribution $n_{\bf k}(t)$ [Eq.~(\ref{quasi_mom})] is shown in the bottom row for the corresponding times. Here $k_{x,y}$ are shown in units of $\pi/a$.}
\label{fig:u_3}
\vspace*{0.4cm}
\end{figure*} 
%
%

\section{Results}
\label{sec:results}

We consider a square optical lattice of size $I\times I$, where $I$ is the odd
number of sites along the $x$ or $y$ direction. Each site is addressed by the
double index $i\equiv (n_x,n_y)$ where $n_x$ and $n_y$ run from  $-(I-1)/2$ to
$(I-1)/2$, and summation over $i$ implies summing over all lattice sites.
We will impose periodic boundary conditions and define the first Brillouin zone such that the quasimomenta $k_x,k_y\in [0,2\pi/a]$, where $a$ is the lattice constant (see also Figs.~\ref{fig:u_3} and \ref{fig:u_7}).
The filling factor is $\rho=N/I^2$, where $N$ is the total particle number, 
\beq
N=\sum_{i}N^0_i(t)+N^{\varphi}_i(t)=\sum_i\left[|a_i(t)|^2+{i}\text{F}_{ii}^G(t,t)-\frac{1}{2}  \right] \,.
\label{total_particle}
\eq
In Eq. \eqref{total_particle}, $N_i^0(t)$ is the number of condensed
particles, while $N^{\varphi}_i(t)$ is the particle number in the 
incoherent cloud, or fluctuation particle number. The total particle 
number $N$ is conserved. This is obeyed by our 
conserving approximations (the BHF and the second-order self-consistent
approximations)\cite{Mauro2015}. We have also confirmed it in our numerics. 
We study the expansion dynamics in dependence on the interparticle interaction $U$, expressed in dimensionless units as
\beq
u=\frac{UN}{J} 
\eq
and express the time in terms of the dimensionless variable $Jt$.

As the initial condition, we assume that at time $t=0$ all particles are condensed in the central site $(0,0)$, that is,
 \bea
a_i(0)&=&\sqrt{\rho I^2}\delta_{i0}, \nn \\
\text{F}_{ij}^G(0,0)&=&-\frac{{i}}{2}\delta_{ij}, \nn \\
\text{F}_{ij}^F(0,0)&=&0. 
\ea
The temporal expansion of the bosonic gas is then expressed in terms of the time-dependent site occupations $N_i(t)=N^0_i(t)+N^{\varphi}_i(t)$, the local fluctuation numbers 
 $N^{\varphi}_i(t)$, and $n_{\bf k}(t)$, the quasimomentum distribution of the atoms spreading over the optical lattice,
\beq
n_{\bf k}(t)=\frac{1}{I^2}\sum_{ij}e^{{i}{\bf k}({\bf r}_i-{\bf
    r}_j)} \langle \hat b_i^{\dagger}(t)\hat b_j(t) \rangle\,.
\label{quasi_mom}
\eq

\subsection{Results without interaction-induced damping}
\label{sec:resultsBHF}

We begin by numerically solving Eqs.~\eqref{HF_eq} which comprise the 
BHF approximation. This does not take into account inelastic quasiparticle
damping effects; however, it captures many of the salient features of
the expansion dynamics, namely, occupation-induced shifts of the
single-particle energies, effects of the coherent BEC dynamics including 
selftrapping, and dynamical creation of the incoherent cloud of 
single-particle excitations beyond the Gross-Pitaevskii equation  
and its exchange with the BEC.

\begin{figure*}[t!]
\centering
\includegraphics[width=0.8\textwidth]{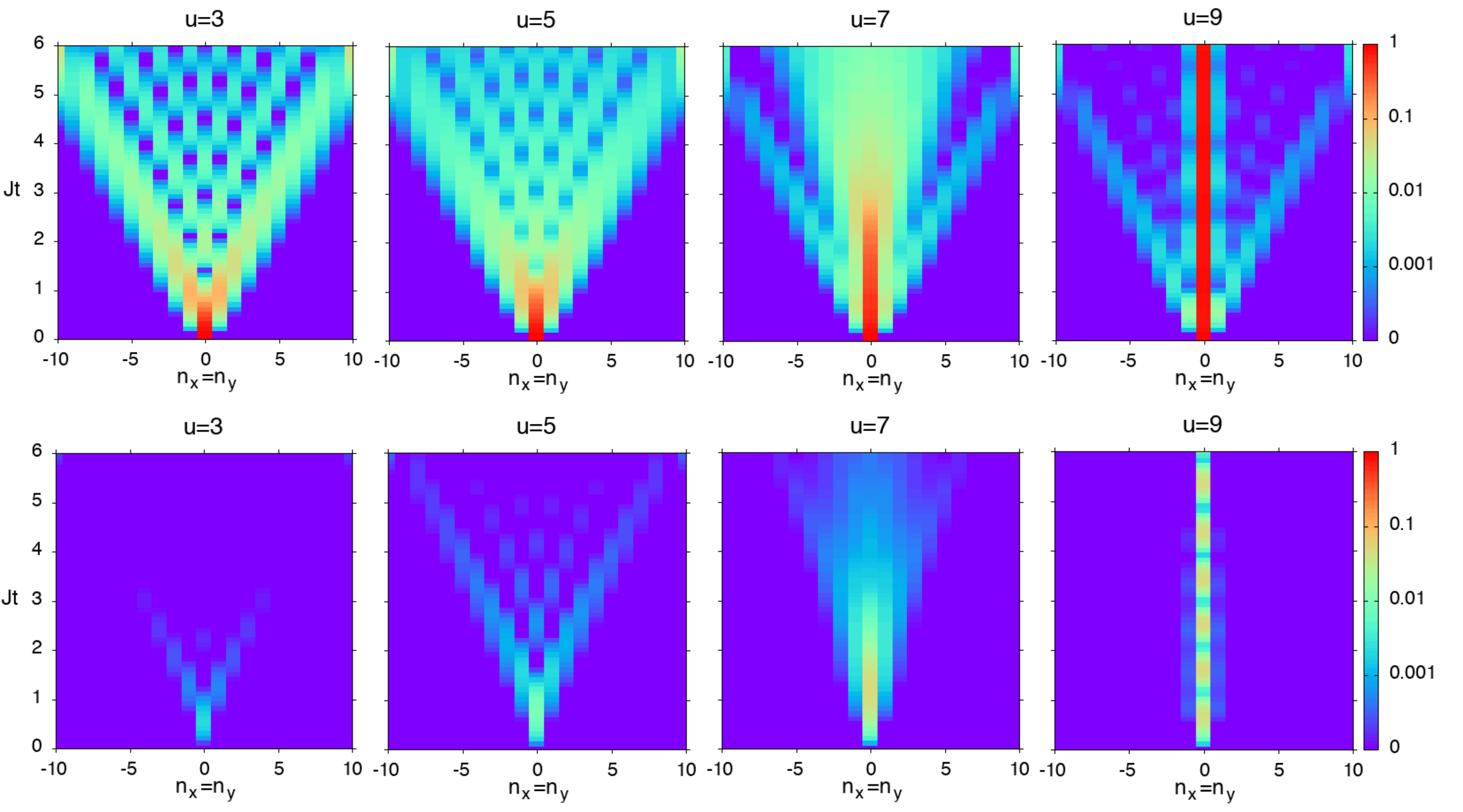}
\caption{(Color online)
Time evolution of the bosonic cloud (top row) and incoherent 
excitations (bottom row) along the lattice diagonal  $n_x=n_y$ with interaction increasing by steps of two from $3$ to $9$.}
\label{fig:diag}
\end{figure*}

\begin{figure}[b!]
\begin{center}
\includegraphics[width=\columnwidth]{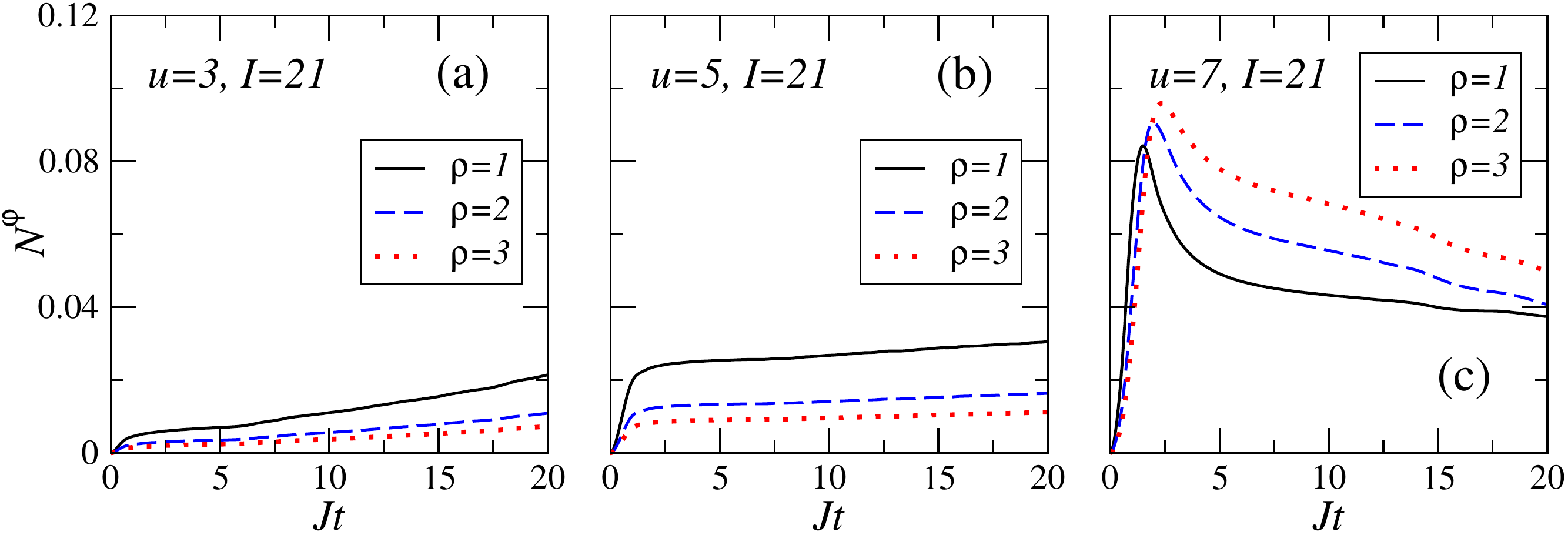}
\end{center}
\vspace*{-0.5cm}
\caption{(Color online)
  Time evolution of the overall quantum fluctuation number
  $N^{\varphi}$  (normalized to the total particle number $N$) for filling
  fractions $\rho=1,\,2,\,3$ and (a) $u=3$, (b) $u=5$, and (c) $u=7$.}
\vspace*{-0.1cm}
\label{fig:fluct_number}
\end{figure}

Figure \ref{fig:u_3} shows the dynamical evolution of the condensate expansion for $I=21$, $\rho=1$, and $u=3$. The top row displays the site occupations $N_i(t)/N$ in real space on a logarithmic scale for different times $Jt$. The expanding cloud reaches the edges of the lattice near $tJ\approx 5$. Thereafter, the local site occupations reappear at the respective, opposite sides of the lattice.
For our symmetric time evolution starting in the center of the lattice, this corresponds to reflection from the boundaries in a finite-size lattice. Since the latter situation is more realistic for experiments, we will refer to this reappearance effect as reflection in the following. The interference patterns appearing for times $Jt\geq 5$ indicate the coherence of the condensate part of the cloud. The fluctuation dynamics is not shown, as the local fluctuations amount to less than 1$\%$ of the total population in this case. The fluctuation dynamics will be discussed in detail in Figs.~\ref{fig:diag} and \ref{fig:fluct_number}.  The bottom row in Fig. \ref{fig:u_3}  presents the lattice-momentum distribution in the first Brillouin zone for the same times. Throughout the time evolution, the momentum distribution is rather broad, showing interference effects of the
condensate again for times $Jt\geq 5$. It indicates a rather homogeneous expansion of the cloud, with velocities given by ${\bf v}_{\bf k}=\partial\varepsilon_{\bf k}/\partial {\bf k}$ and the square lattice dispersion $\varepsilon_{\bf k}=-2J[\cos (k_xa)+\cos(k_ya)]$, as also seen from the real-space pictures.

\begin{figure*}[t!]
\centering
\includegraphics[width=0.99\textwidth]{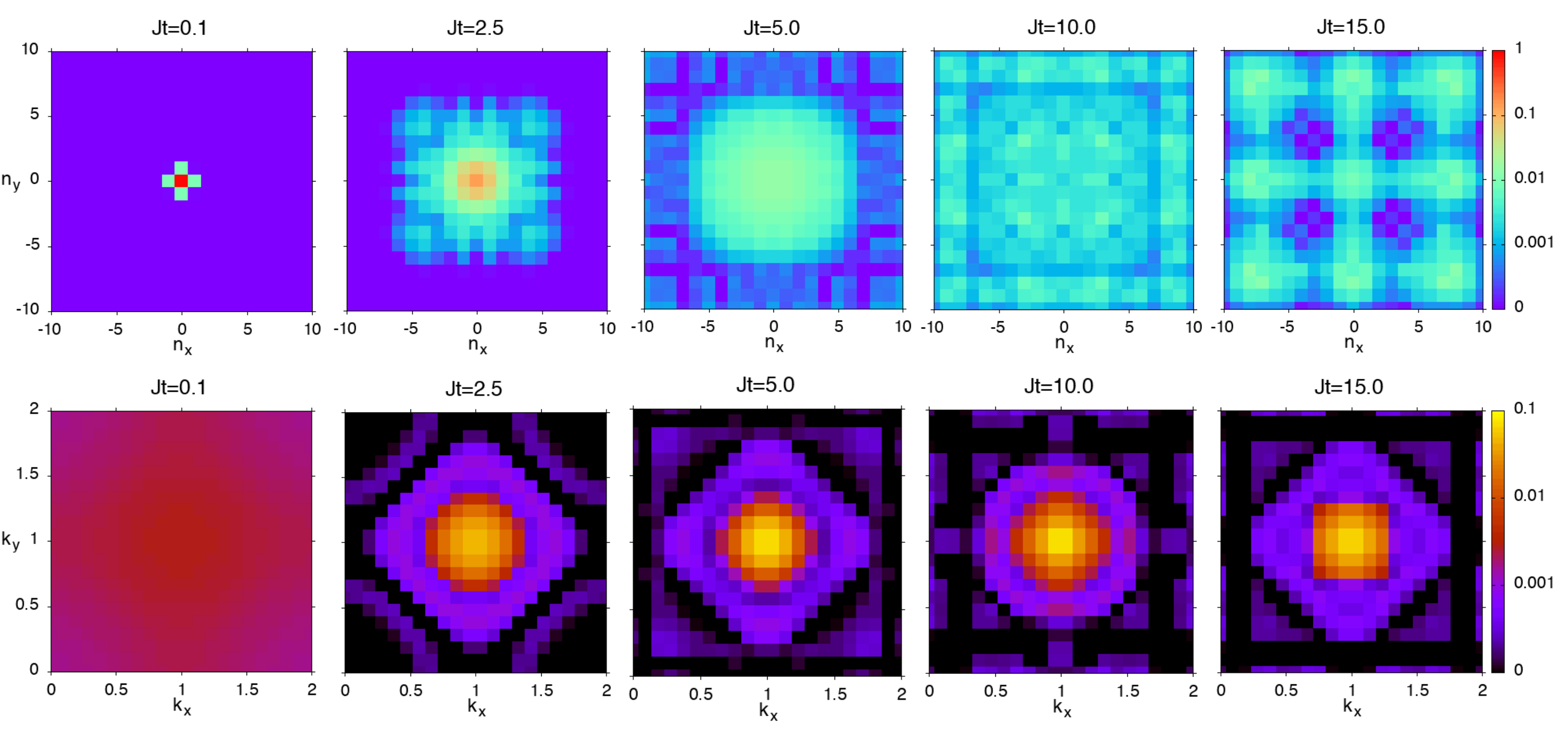}
\caption{(Color online) Expansion dynamics of the BEC cloud for the same parameters as in
  Fig.~\ref{fig:u_3}, but for interaction strength $u=7$.
  Distinct self-trapping effects are visible, both in the spatial density distribution (top row) and in the momentum distribution (bottom row).} 
\label{fig:u_7}
\end{figure*}

The expansion dynamics changes drastically when the interaction is increased,
as shown in Fig.~\ref{fig:u_7} for $u=7$. The real-space pictures (top row
in Fig.~\ref{fig:u_7}) show that the BEC cloud expands more slowly, where the 
cloud separates in a high-density central part and a low-density, 
halolike structure surrounding it, before the particles get reflected from
the boundaries and interference effects set in again. This slow expansion 
is confirmed by the momentum distribution shown in the bottom row of 
Fig.~\ref{fig:u_7}. Throughout the time evolution, it remains strongly
peaked around ${\bf k}=(\pi/a,\pi/a)$ where the group velocity 
${\bf v}_{\bf k}$ of the square lattice vanishes. 
This behavior  is  similar to the experimental observation
\cite{Schneider2013}, where for a noninteracting gas a homogenous square
spread is observed, whereas for finite interaction a bimodal structure of a
slow central cloud surrounded by fast square-shaped background is seen in
two dimensions.

In order to better understand the behavior of the bimodal structure  we plot
in Fig.~\ref{fig:diag} the expansion of the bosonic cloud along the lattice 
diagonal $n_x=n_y$ versus time for different values of the interaction $u$ and
for both the BEC (top row) and the incoherent cloud (bottom row). 
While for interaction strength $u\leq 5$ the BEC spreads essentially
homogeneously with a slight maximum at the expansion front, the bimodal 
expansion of the BEC is clearly seen for $u\geq 7$. 
This can be explained by the self-trapping effect known from 
bosonic Josephson junctions \cite{Smerzi1997,Albiez2005}.  It is due to an
energy mismatch between neighboring sites, that is, it occurs in the
nonlinear Gross-Pitaevskii dynamics when the difference 
between the total condensate energies $u_i=UN^0_i(t)[N^0_i(t)-1]/2J$
on neighboring sites $i$ and $j$ or 
neighboring potential wells exceeds a critical value
\cite{Smerzi1997,Mauro2009}. This is fulfilled at the boundaries of
the central BEC cloud for $u\geq 7$. As a result, the tunneling  
of condensate particles from the central, high-density sites to the outer,
low-density sites is inhibited, leading to a reduced expansion speed of the
central cloud, as seen in the top row of Fig.~\ref{fig:diag}. By contrast, 
particles on the low-density sites within the halo are not subject to 
self-trapping and therefore propagate outward ballistically with constant,
high speed, in agreement with experiment \cite{Schneider2013}. 
In fact, as seen from Fig.~\ref{fig:diag}, the halo speed in the $x$ direction 
reaches the Lieb-Robinson limit \cite{Lieb1972}, which for our tight-binding 
square lattice is $v_{x}^{\text{max}}=|{\bf v}_{{\bf k}=(\pi/2a,0)}|=2Ja$. 
As a result of the different expansion
speeds, the central cloud and the halo become spatially separated,
leading to a distinct forerunner. This explanation of the bimodal 
expansion is corroborated by the fact that it is observed only in the 
expansion of the coherent BEC (top row of Fig.~\ref{fig:diag}),
not in the incoherent cloud
(bottom row of Fig.~\ref{fig:diag}), as expected from the Gross-Pitaevskii
dynamics. Rather, the incoherent excitations are dragged along by 
the coherent, central cloud and their number is imperceptibly small in 
the halo for $u\geq 7$. As a remarkable observation, there seems 
to be an interesting interplay between nonlinear self-trapping and 
interference for long times. 
After reflection from the lattice boundaries, the local BEC amplitude 
may constructively interfere so that locally the selftrapping threshold
is reached again, leading to multiple quasilocalized BEC clouds, seen 
in Fig.~\ref{fig:u_7}, top right panel ($Jt=15$), as the nine 
high-occupation regions. They cannot be a mere interference effect,
as they are not observed in Fig.~\ref{fig:u_3} for small interaction ($u=3$).


The role of the incoherent cloud in the expansion dynamics can be 
analyzed by monitoring the time evolution of the total fluctuation number 
and the energy content of the incoherent cloud and the BEC, respectively.
Figure~\ref{fig:fluct_number} shows the noncondensate or fluctuation 
fraction $N^{\varphi}(t)$
(i.e., the fluctuation number normalized to the total particle number)
for various interaction strengths $u$ and filling fractions $\rho$. 
First, we observe that $N^{\varphi}(t)$ remains in the range of a few percent 
throughout the time evolution for all parameter values of the `$21\times 21$ 
lattice. For all parameter values, there is a steep, initial rise of the 
fluctuation fraction within the time of the first tunneling process of the 
time evolution, $t\approx 1/J$. This is presumably due to the abrupt change 
of the occupation numbers on the lattice sites neighboring the central one 
and the consecutive strong perturbation of the BEC. 
Thereafter, for weak interactions ($u=3,\,5$) the fluctuation fraction 
increases monotonically with an approximately constant slope. 
In this weak-interaction regime, $N^{\varphi}(t)$ is 
approximately inversely proportional to the filling fraction $\rho$, as 
expected from the perturbative expansion about a homogeneous 
(position-independent) Gross-Pitaevskii saddle point: The BEC density creates
a confining potential for the fluctuations, so increasing density
suppresses fluctuations. 
However, when self-trapping effects set in for strong interaction ($u=7$),
the behavior after the initial, strong increase gets reversed. After a
pronounced maximum, the total fluctuation density decreases again with time
and settles to a reduced, constant value. 
It seems that the self-trapping-induced, slowed-down dynamics does not 
support the initial, enhanced fluctuation density, so incoherent 
excitations recondense into the BEC. In this case, the fluctuation fraction 
$N^{\varphi}$ also increases with the filling fraction $\rho$.


\begin{figure}[t!]
\begin{center}
\includegraphics[width=0.95\columnwidth]{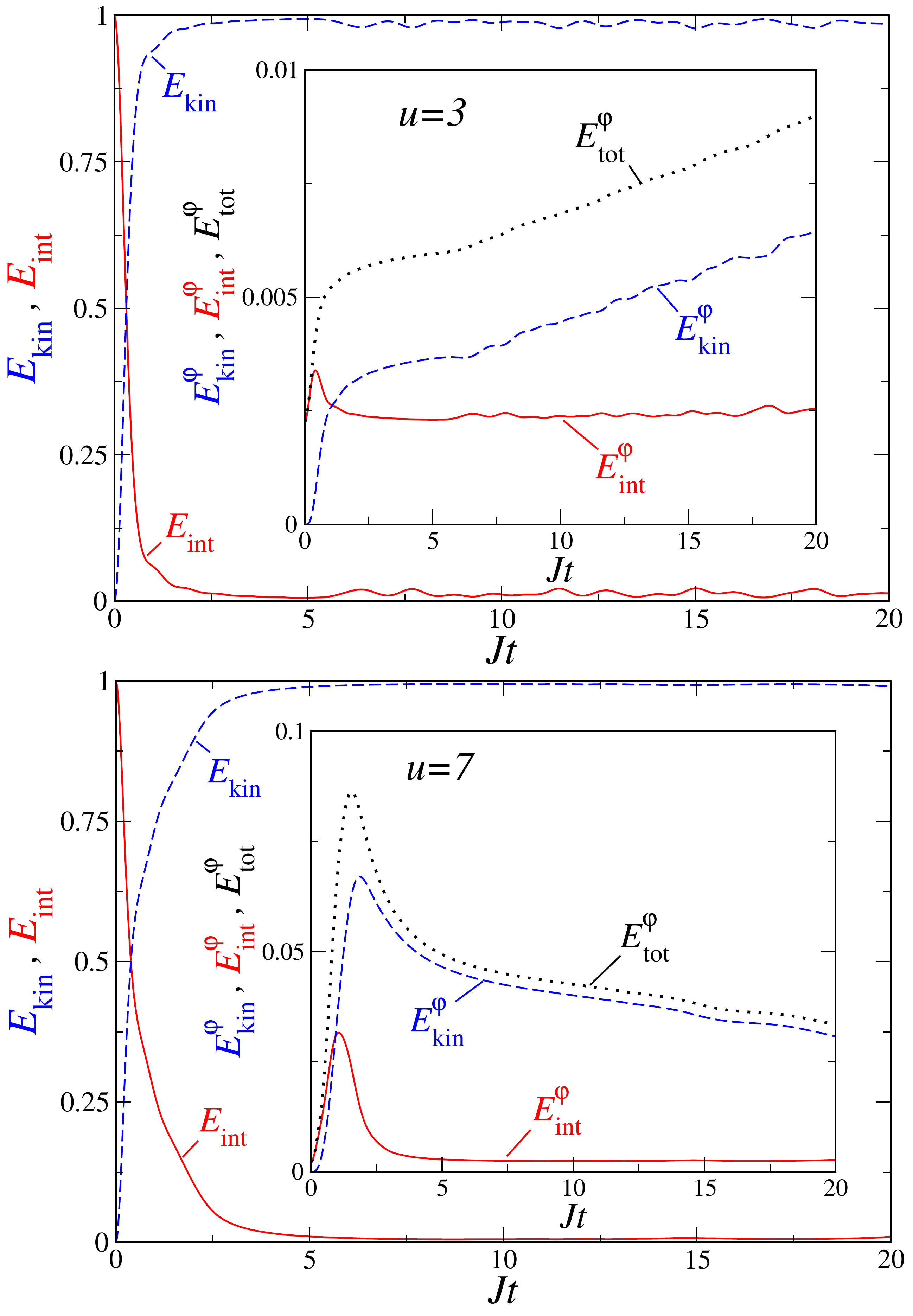} 
\end{center}
\vspace*{-0.3cm}
\caption{(Color online) Time evolution of the mean kinetic energy (dashed lines)
  and interaction energy (solid lines)
  [Eqs.~\eqref{eq:kin_energy} and \eqref{eq:int_energy}], normalized to the total
  energy $E_0=u(N-1)/2$ of the atomic gas loaded on the $21\times 21$ 
  lattice, with $\rho=1$ for interaction parameters $u=3$ (top panel)  
  and $u=7$ (bottom panel). 
  The insets show the kinetic and interaction energies of
  the quantum fluctuations together with their total energy 
  $E_{tot}^{\varphi}(t)=E_{kin}^{\varphi}(t)+E_{int}^{\varphi}(t)$ (dotted line).}
\label{fig:energy}
\end{figure}

The time evolution of the kinetic energy  $E_{kin}(t)$ and the interaction
energy of the total system $E_{int}(t)$ and the respective contributions to the
fluctuation energy are shown in Fig.~\ref{fig:energy}. 
The corresponding energy expectation values are computed 
from the on-site and nearest-neighbor propagators, respectively, as 
 \bea
E_{kin}(t)&=&E_{kin}^0(t)+E_{kin}^{\varphi}(t)  \label{eq:kin_energy} \\
&=&-\frac{{i}J}{2}\sum_{\langle i,j\rangle}\,
\text{Tr}\left[{\bf C}_{ij} + {\bf F}_{ij}\right] \nn \\
E_{int}(t)&=&E_{int}^0(t)+E_{int}^{\varphi}(t)  \label{eq:int_energy}\\
&=&\frac{{i}}{4}\sum_i \, \text{Tr}\left[ {\bf S}^{HF}_i{\bf C}_{ii}+{\bf \Sigma}^{HF}_i{\bf F}_{ii} \right], \nn 
\ea 
where the superscripts $0$ and $\varphi$ refer to the contributions from
condensed atoms and quantum fluctuations, respectively, and $\langle
i,j\rangle$ indicates summation over nearest-neighbor sites.
While our conserving approximation preserves the total energy conservation,
the initially purely interaction energy is quickly transformed into kinetic 
energy as the central site is depopulated (main panels in
Fig.~\ref{fig:energy}). Note that, counterintuitively, this energy transfer 
is slowed down for strong repulsion, $u=7$, because the depopulation of 
highly occupied sites is slowed down by selftrapping effects (see also 
Fig.~\ref{fig:diag}). 

The time evolution of the fluctuation energies (insets of
Fig.~\ref{fig:energy}) reflects the fluctuation-number dynamics discussed in 
Fig.~\ref{fig:fluct_number}. For small interaction ($u=3$), after the initial, 
steep increase, the kinetic fluctuation energy $E_{kin}^{\varphi}(t)$
continues to increase linearly in accordance with the fluctuation number 
[Fig.~\ref{fig:fluct_number} (a)], while the interaction energy 
$E_{int}^{\varphi}(t)$ settles to a constant value. This shows that, 
although the overall fluctuation number increases with the volume of the 
cloud, the fluctuation number on each site remains on a constant, small level.
On-site two-body interactions among the fluctuations are therefore expected to
be small. 
For strong interaction in the self-trapping regime ($u=7$), the 
fluctuation kinetic energy follows the evolution of the total fluctuation
number as well [Fig.~\ref{fig:fluct_number} (c)], while the interaction
energy settles again to a constant value after a pronounced initial peak.


\subsection{Small lattices with damping}
\label{sec:coll}

\begin{figure}[t!]
\begin{center}
\includegraphics[width=\columnwidth]{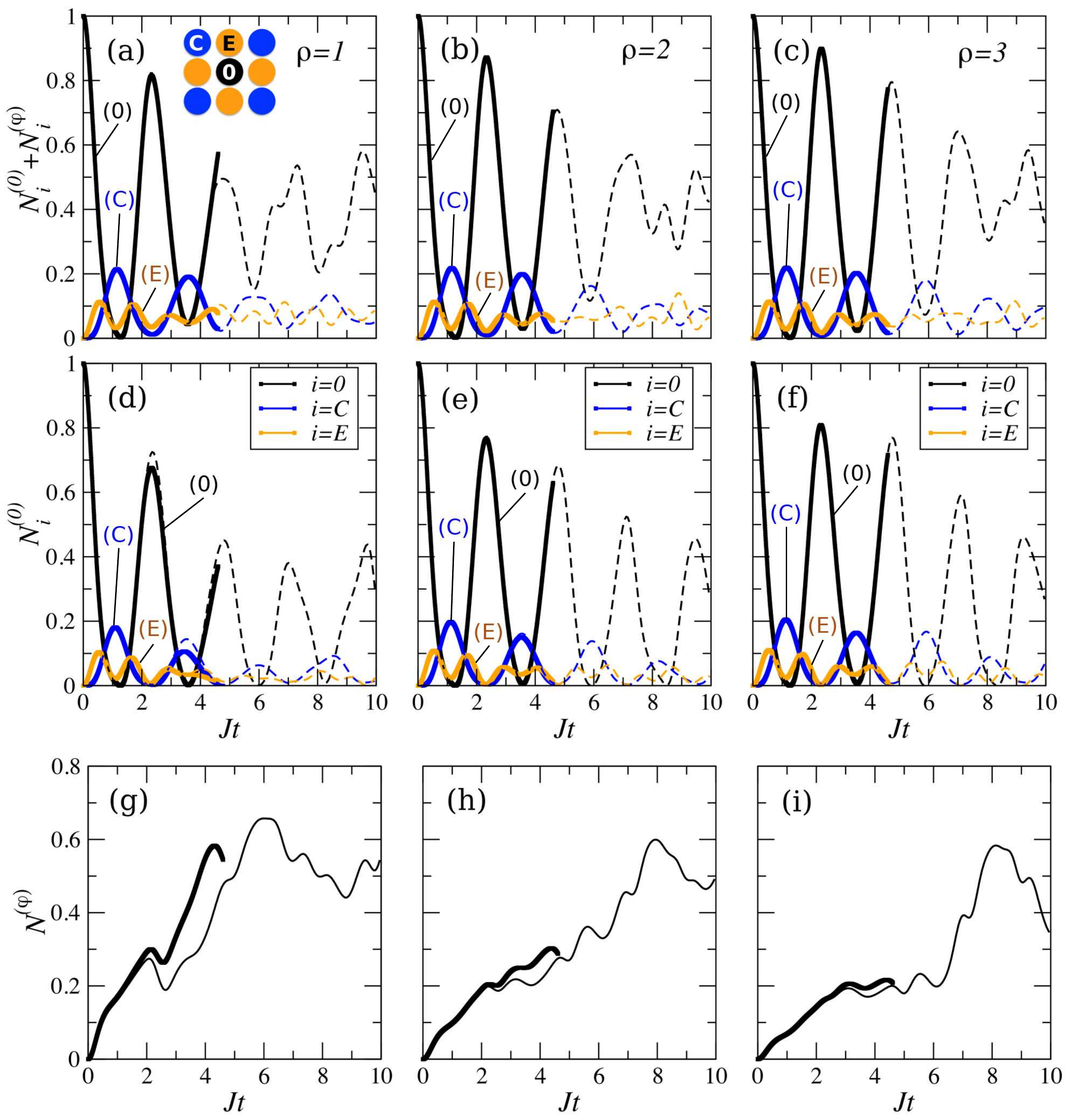}
\end{center}
\caption{(Color online)
  Site-resolved expansion in a $3\times 3$ lattice for $u=3$ 
  and different
  filling factors $\rho=1$, $2$, and $3$. Here $N_i^{(0)}(t)$ denotes the
  condensate number, $N_i^{(\varphi)}(t)$ the fluctuation number on site $i$,
  and $N^{(\varphi)}(t)$ the total number of fluctuations, each normalized to 
  the total particle number $N$.
  Black, blue (dark gray), and orange (light grey) lines represent the
  central (0), corner (C), and edge sites (E), respectively, as indicated
  and visualized
  by the color-coded lattice shown in the inset. Thick lines represent the
  solutions of the self-consistent second-order approximation including
  inelastic collisions; thin or dashed dashed lines correspond to the solutions
  obtained within the BHF approximation in all panels.}
\label{fig:damp_u3}
\end{figure}

\begin{figure}[t!]
\begin{center}
\vspace*{-0.2cm}
\includegraphics[width=\columnwidth]{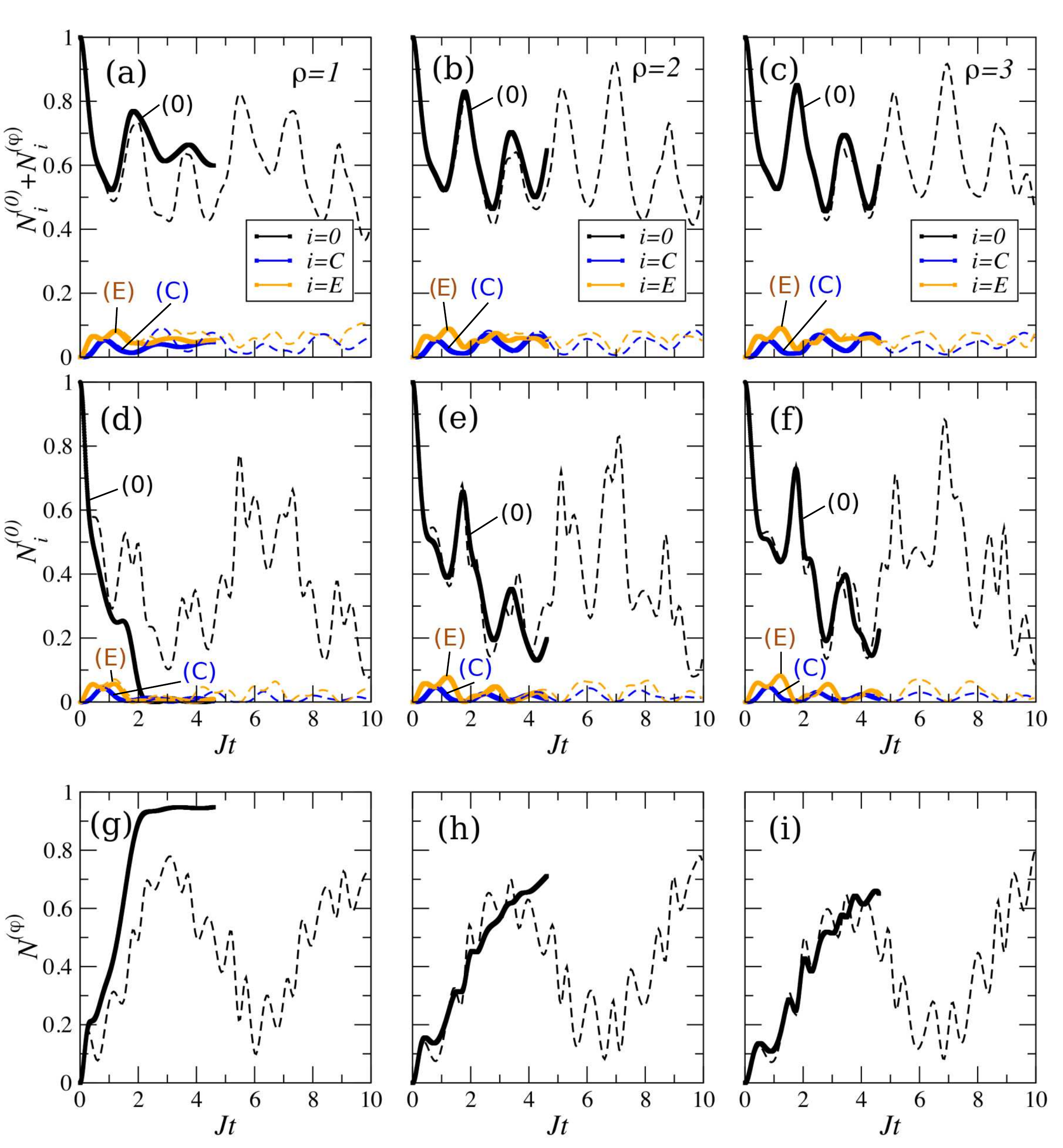}
\end{center}
\vspace*{-0.12cm}
\caption{(Color online) Site-resolved expansion in a $3\times 3$ lattice as in
  Fig.~\ref{fig:damp_u3}, but for interaction strength $u=7$.}
\label{fig:damp_u7}
\end{figure}

We now investigate the influence of collisions  of the incoherent 
excitations and damping by solving the second-order self-consistent
approximation described by Eqs. \eqref{eq:spectral1}--\eqref{eq:cond}. 
This requires time evolving the propagators and 
self-energies for two different time arguments. Some details about how to deal 
with convolution integrals in this case are given in Appendix B.
For reasons of numerical costs of 
computing the second-order self-energies we only consider a $3\times 3$ 
lattice and limited evolution time here. The corresponding results are 
shown in Figs.~\ref{fig:damp_u3} and \ref{fig:damp_u7} for all the 
different inequivalent lattice sites and compared with the BHF results (dashed lines).

First, we note that in the small lattice the fluctuation fraction is 
generally higher than in the large $21\times 21$ lattice. This may be traced
back to the fact that for our initial conditions in the $3\times 3$ lattice  
the initial condensate number on the central site is smaller than in the 
$21\times 21$ lattice and thus fluctuations are less strongly suppressed.
For the same reason, self-trapping effects are weaker.
Second, for $u=3$ the time evolution including collisions of incoherent 
excitations reproduces the BHF approximation for all considered 
quantities and filling fractions quantitatively rather well 
(see Fig.~\ref{fig:damp_u3}), as already conjectured at the end of 
Sec.~\ref{sec:resultsBHF}. 
The on-site condensate occupation numbers $N_i^{0}(t)$ perform pronounced,
weakly damped Josephson-like oscillations \cite{Mauro2009}.
Finally, for $u=7$ the total local occupation numbers $N_i^0(t)+N_i^{\varphi}(t)$ 
agree again quantitatively well with the BHF approximation, as seen from 
Figs.~\ref{fig:damp_u7}(a)--\ref{fig:damp_u7}(c).
However, in this case and for $\rho=1$, the fraction of fluctuations 
produced by strong interaction and second-order collisions, 
deviates strongly from the prediction of the BHF approximation. 
This leads to fast condensate depletion on all sites, and condensate
oscillations are quickly damped, as seen in Fig.~\ref{fig:damp_u7}(d) and \ref{fig:damp_u7}(g). On the other hand, as the filling 
fraction is increased to $\rho=2$ and $3$, the agreement with the
BHF approximation is restored for propagation times up to $Jt\approx 4$
[see Fig.~\ref{fig:damp_u7}(e), \ref{fig:damp_u7}(f), \ref{fig:damp_u7}(h),
and \ref{fig:damp_u7}(i)]. Note that these filling
fractions correspond to initial populations of the central site of 
18 and 27, respectively, which are still far below the initial central-site
occupation of $21^2=441$ for the $21\times 21$ lattice at $\rho =1$.
Considering that the importance of fluctuations (and therefore their 
interactions) decreases with increasing condensate population, 
we conclude that the BHF results of Sec.~\ref{sec:resultsBHF} for 
a $21\times 21$ lattice should be reliable at least for the initial 
time evolution up to $Jt\approx 5$, even though quasiparticle collisions are
not taken into account there. Consequently, the bimodal expansion separated
into a strongly populated central cloud and a weak quantum coherent 
halo should be observable in high-density clouds, similar to the 
experimental findings of Ref.~\cite{Schneider2013}. 
We note that for the discrete density of states of the finite-size
lattice, a homogeneous 2D BEC would be thermodynamically stable in equilibrium 
as long as the particle density does not drop below a critical value 
$n_{crit}(T)$. Estimating the final-state equilibrium temperature $T_{\infty}$ 
from the interaction energy per particle of the initial state, we have 
for our systems $n_{crit}(T_{\infty}) = {\it O}(10^{-1})$, significantly below
the average particle density of 1 per site. Therefore, the condensate
depletion is due to the nonequilibrium dynamics rather than a thermodynamical
instability.


\section{Conclusion}
\label{sec:conclusion}

We have studied the temporal expansion dynamics of an ultracold  Bose gas 
in the two-dimensional Bose-Hubbard model, where initially all bosons 
are trapped and condensed at the central site of a square lattice. 
Within our formalism we
were able to analyze separately (i) the semiclassical Gross-Pitaevskii 
dynamics of the BEC, (ii) the dynamics of quantum fluctuations and
(iii) their inelastic two-body interactions as well as the mutual influence 
of these components on each other. After the hopping is switched 
on, the bosons spread over the lattice in a nontrivial way. We showed 
that the expansion dynamics depends crucially on interactions. One can 
clearly distinguish a strongly interacting regime (for our system, $u\geq7$) 
when the condensate cloud effectively splits into two parts: a slowly 
expanding high-density part at the
center of the lattice and a fast, quantum coherent, low-density halo or
forerunner at the rim of the cloud. This bimodal structure is in qualitative 
agreement with the experiment \cite{Schneider2013} and is interpreted as due
to nonlinear selftrapping effects, since doublon dynamics should 
not play a role in this high-density regime. The existence of a coherent 
forerunner with average particle number per site of about 1 for strong 
interactions $u=7,\,9$ (see Fig.~\ref{fig:diag}, top panels) shows that a
quantum distillation effect \cite{Weiss2015} also exists for high initial 
particle density.
We obtained constant expansion speed, indicative of ballistic motion,
for the front of both the BEC and the incoherent cloud 
(see Fig.~\ref{fig:diag}), as one expects for a coherent wave function. 
While for the incoherent cloud one might expect diffusive expansion as for 
classical particles on a lattice, we attribute its linear-in-time expansion
to a dragging effect of the incoherent cloud by the BEC.
Furthermore, we demonstrated that the velocity of the forerunner is bounded 
from above and reaches the maximum group velocity of the system, 
which represents the Lieb-Robinson limit \cite{Lieb1972} for this case. 

We also showed that in high-density, interacting clouds the quench dynamics 
leads to an initial, explosive generation of fluctuations, albeit with a 
small overall amplitude. As the cloud further expands, for weak interaction 
($u=3$, $5$) the fluctuation number continues to increase, but linearly in
time with a moderate slope. For stronger interaction ($u=7$), 
the role of the condensate as a confining potential for the fluctuations
starts to dominate, so the initially created fluctuation numbers
are no longer supported by the further expansion dynamics, leading to 
a peaklike time evolution of the fluctuation number. In this way, the 
fluctuations remain confined to a low number in high-density systems
for all interaction parameters. 

For smaller lattices we studied damping effects due to quasiparticle 
collisions. Since it is numerically a challenging problem, we
reduced the analysis to $3\times 3$ lattices. In this case, the fluctuation 
numbers are much higher than in larger lattices, as expected. In low-density 
systems, the collision-induced damping can lead to a fast depletion of the 
condensate population, transforming the system into an incoherent gas. 
In the long-time limit, this should lead to a thermal gas \cite{Anna2016} 
above the condensation temperature.
For larger filling fractions (or initial central site occupation), however, 
the coherent BEC dynamics is recovered for significant evolution times.
Extrapolating this finding to the large lattice systems, we concluded that
the bimodal expansion should be robust against collisions.
This analysis shows the possibly important role of fluctuations, 
depending on the lattice size and interaction also noted in the experiment 
\cite{Schneider2013}. In addition, it provides an explanation 
why the bimodal expansion was experimentally observed 
\cite{Schneider2013}, despite the ubiquitous presence of fluctuations.
In the future it will be interesting to study collision effects for 
disordered as well as periodically driven Bose-Hubbard lattices.

\vspace*{0.5cm}

\acknowledgments
We gratefully acknowledge useful discussions with Ulrich Schneider.
This work was financially supported by the Deutsche Forschungsgemeinschaft
within the Cooperative Research Center SFB/TR 185 (Grant No. 277625399)
and the Cluster of Excellence ML4Q (Grant No. 390534769).

\appendix

\onecolumngrid

\section{Symmetry relations}
\label{symmetry_rel}

Our equations simplify if we make use of the symmetry relations listed below. For the components of the spectral function and statistical functions the following equations hold:
\begin{equation}
\begin{aligned}
\text{A}_{ij}^{\ov G}(t,t')&=-\text{A}_{ij}^{G}(t,t')^*=-\text{A}_{ji}^{G}(t',t) \\ 
\text{A}_{ij}^{\ov F}(t,t')&=-\text{A}_{ij}^{F}(t,t')^*=\text{A}_{ji}^{F}(t',t)^* \\ 
\text{F}_{ij}^{\ov G}(t,t')&=-\text{F}_{ij}^{G}(t,t')^*=\text{F}_{ji}^{G}(t',t) \\ 
\text{F}_{ij}^{\ov F}(t,t')&=-\text{F}_{ij}^{F}(t,t')^*=-\text{F}_{ji}^{F}(t',t)^*.
\end{aligned}
\label{symmetry}
\end{equation}
The selfenergies which enter Eqs.~\eqref{general_eq1}--\eqref{general_eq3}
also obey the symmetry relations 
\begin{equation}
\begin{aligned}
\gamma_{ij}^G(t,t')^*&=-\gamma_{ij}^{\ov G}(t,t')=\gamma_{ji}^G(t',t) \\
\gamma_{ij}^F(t,t')^*&=-\gamma_{ij}^{\ov F}(t,t')=\gamma_{ji}^{\ov F}(t',t) \\
\Gamma_{ij}^G(t,t')^*&=-\Gamma_{ij}^{\ov G}(t,t')=\Gamma_{ji}^G(t',t) \\
\Gamma_{ij}^F(t,t')^*&=-\Gamma_{ij}^{\ov F}(t,t')=\Gamma_{ji}^{\ov F}(t',t) \\
\Pi_{ij}^G(t,t')^*&=-\Pi_{ij}^{\ov G}(t,t')=-\Pi_{ji}^G(t',t) \\
\Pi_{ij}^F(t,t')^*&=-\Pi_{ij}^{\ov F}(t,t')=-\Pi_{ji}^{\ov F}(t',t)\,.
\end{aligned}
\end{equation}

\section{Convolution integrals in equations of motion}
\label{integrals}

We can use the symmetry relations listed in Appendix \ref{symmetry_rel} also
in the convolution integrals which enter our equations of motion
\eqref{eq:spectral1}--\eqref{eq:stat2}. Consider, for example, the integrals
in Eq.~\eqref{eq:stat1},
\begin{equation}
\begin{aligned}
-{i}\sum_k\int \limits_{0}^{t}d \ov{t}[\Gamma^G_{ik}(t,\ov{t})\text{F}^G_{kj}(\ov{t},t')+\Gamma^F_{ik}(t,\ov{t})\text{F}^{\ov{F}}_{kj}(\ov{t},t')]  
+{i}\sum_k\int \limits_{0}^{t'}d \ov{t}[\Pi^G_{ik}(t,\ov{t})\text{A}^G_{kj}(\ov{t},t')+\Pi^F_{ik}(t,\ov{t})\text{A}^{\ov{F}}_{kj}(\ov{t},t')]. 
\end{aligned}
\label{int_1}
\end{equation}
The symmetry relations allow us to split the interval of integration in such a
way that we can rewrite the integrals with the arguments corresponding to the
later time as first arguments. Hence, we get, for the integral \eqref{int_1}, 
\begin{equation}
\begin{aligned}
\phantom{+}{i}\sum_k\int \limits_{0}^{t'}d \ov{t}[\Gamma^G_{ik}(t,\ov{t})\text{F}^G_{jk}(t',\ov{t})^*+\Gamma^F_{ik}(t,\ov{t})\text{F}^{F}_{jk}(t',\ov{t})^*]
&-{i}\sum_k\int \limits_{t'}^{t}d \ov{t}[\Gamma^G_{ik}(t,\ov{t})\text{F}^G_{kj}(\ov{t},t')-\Gamma^F_{ik}(t,\ov{t})\text{F}^{F}_{kj}(\ov{t},t')^*] \\
&+{i}\sum_k\int \limits_{0}^{t'}d \ov{t}[\Pi^G_{ik}(t,\ov{t})\text{A}^G_{jk}(t',\ov{t})^*+\Pi^F_{ik}(t,\ov{t})\text{A}^{F}_{jk}(t',\ov{t})]. 
\end{aligned}
\end{equation}
We proceed in an analogous way for the other integrals of
Eqs.~\eqref{eq:spectral1}--\eqref{eq:stat2} and then solve the final system of equations numerically.  

\vspace*{-0.3cm}

\twocolumngrid

\end{document}